\documentclass[runningheads]{llncs}
\usepackage[dvipdfmx]{graphicx}
\usepackage[dvipdfmx]{color}
\usepackage[table,xcdraw]{xcolor}
\usepackage{subfig} 
\usepackage{adjustbox}
\usepackage{multirow}
\usepackage[subtle,margins=normal,leading=normal]{savetrees} 
\newcommand*\samethanks[1][\value{footnote}]{\footnotemark[#1]}
\begin{document}
\title{Automated Pancreas Segmentation Using Multi-institutional Collaborative Deep Learning}
%
%
\author{Pochuan Wang\thanks{equal contribution}\inst{1} \and
Chen Shen\samethanks[1]\inst{2} \and
Holger R. Roth \inst{3} \and \\
Dong Yang \inst{3} \and
Daguang Xu \inst{3} \and
Masahiro Oda \inst{2} \and
Kazunari Misawa \inst{4} \and
Po-Ting Chen \inst{5} \and \\
Kao-Lang Liu \inst{5} \and
Wei-Chih Liao \inst{5} \and
Weichung Wang \inst{1} \and
Kensaku Mori \inst{2}
}
\authorrunning{P. Wang et al.}
\titlerunning{Automated Pancreas Segmentation Using Collaborative Deep Learning}
%
\institute{National Taiwan University, Taiwan \and
Nagoya University, Japan \and
NVIDIA Corporation, United States \and
Aichi Cancer Center, Japan \and
National Taiwan University Hospital, Taiwan
}
\maketitle              
\begin{abstract}
The performance of deep learning based methods strongly relies on the number of datasets used for training. Many efforts have been made to increase the data in the medical image analysis field. However, unlike photography images, it is hard to generate centralized databases to collect medical images because of numerous technical, legal, and privacy issues. In this work, we study the use of federated learning between two institutions in a real-world setting to collaboratively train a model without sharing the raw data across national boundaries. We quantitatively compare the segmentation models obtained with federated learning and local training alone. Our experimental results show that federated learning models have higher generalizability than standalone training. 

\keywords{Federated Learning \and Pancreas Segmentation \and Neural Architecture Search.}
\end{abstract}
\section{Introduction}
Recently, deep neural networks (DNNs) based methods have been widely utilized for medical imaging research. High-performing models that are clinically useful always require vast, varied, and high-quality datasets. However, it is expensive to collect a large number of datasets, especially in the medical field. Only well-trained experts can generate acceptable annotations for DNN training, making annotated medical images even more scarce. Furthermore, medical images from a single institution can be biased towards specific pathologies, equipment, acquisition protocols, and patient populations. The low generalizability of DNNs models trained on insufficient datasets is critical when applying deep learning methods for clinical usages.

To improve the robustness with scant data, fine-tuning is an alternative way to learn the knowledge from pre-trained DNNs. The fine-tuning technique starts training from a pre-trained weight instead of random initialization, which has been proved helpful in medical image analysis \cite{shin2016deep,Nima2016-tn}, which exceeds the performance on training a DNN from scratch. However, fine-tuned models can still have high deficiencies in generalizability \cite{Chang20180-kc}. When a model is pre-trained on one data (source data) and then fine-tuned on another data (target data), the trained model tends to fit on target data but lose the representation on source data \cite{li2017learning}.

Federated learning (FL) \cite{McMahan2017-hb} is an innovation for solving this issue. It can collaboratively train the DNNs using the datasets from multiple institutions without creating a centralized dataset \cite{Li2019-wq,Sheller2019-mj}. Each institution (client) trains with local data using the same network architecture decided in advance. After a certain amount of local training, each institution regularly sends the trained model to the server. The server only centralizes the weights of the model to aggregate, and then send them back to each client.

In this work, we collaboratively generated and evaluated an FL model for pancreas segmentation without sharing the data. Our data consists of healthy and unhealthy pancreas collected at the two institutions from different countries (Taiwan and Japan). Throughout this study, we utilized the model from coarse-to-fine network architecture search (C2FNAS)~\cite{Yu2019-qi} with an additional variational auto-encoder (VAE)~\cite{Myronenko2019-ye} branch to the encoder endpoint. FL dramatically improved the generalizability of models on server-side and client-side for both datasets. To the best of our knowledge, this is the first time performing FL for building a pancreas segmentation model from data hosted at multi-national sites.
\section{Methods}

\subsection{Federated Learning}
\begin{figure}
\centering
\includegraphics[width=0.9\textwidth]{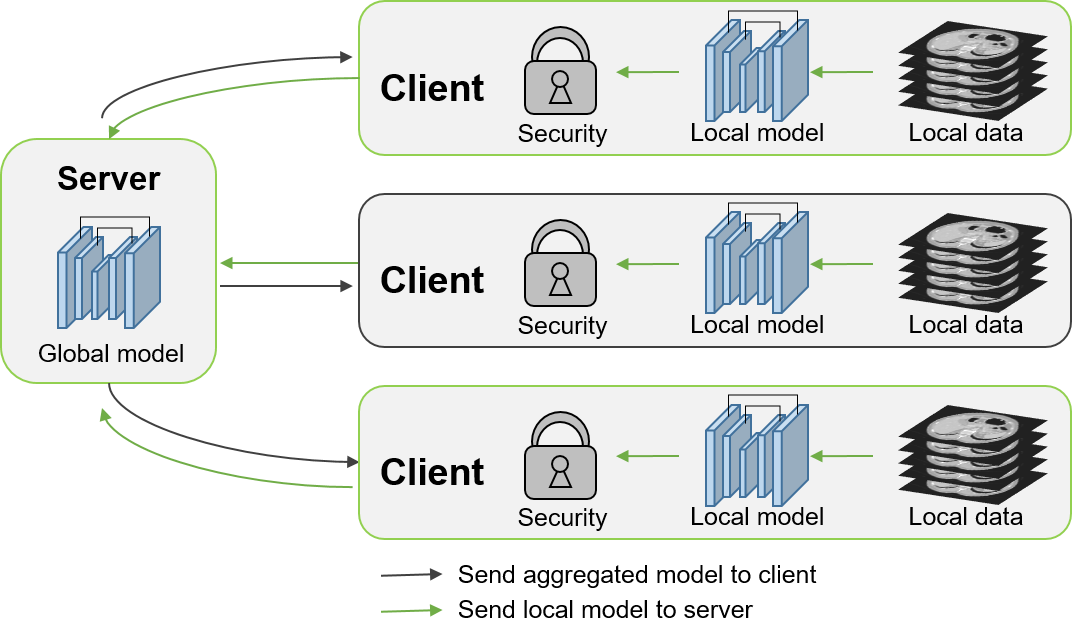}
\caption{The architecture of federated learning system.} 
\label{fig:fd_arch}
\end{figure}
FL can be categorized into different types based on the distribution characteristics of data \cite{Yang2019-qiang}. In this work, we only focus on horizontal architecture, which is illustrated in Fig. \ref{fig:fd_arch}. This type of FL allows us to train with datasets from different samples distributed across clients. 

A horizontal FL system consists of two parts: {\it server} and {\it clients}. 
The server manages the training process and generates a {\it global model}, and the client train with local data to produces a {\it local model}. The server receives trained weights from each client and aggregates them into a global model. The clients train with the local dataset and send the weights to the server. We call the process of generating one global model one round.

The workflow consists of the following steps:
\begin{enumerate}
  \item Start the server. The server-side sets the gPRC communication ports, SSL certificate, the maximum and minimum numbers of clients. 
  \item Start the client. Use client-side configuration to initialize. Then use the credential to make a login request to the server.
  \item Client-side downloads the current global model from the server and fine-tuning the model with the local dataset. Then, only submit the model to the server and wait for other clients.
  \item Once the server receives the model from a previously defined minimum number of clients, it will aggregate them into a new global model. 
  \item The server updates the global model and finishes one round.
  \item Go back to 3. for another round.
\end{enumerate}

The model shared among the server and clients is only weight parameters, protecting the privacy for local data. To build the server-client trust, the server-side uses token throughout the process. SSL certificate authority and gPRC communication ports were adopted to improve security.

\subsection{Data Collection}
We use two physically separated clients in this work in order to try FL in the real-world setting. Two different datasets from two institutions from two different countries were applied.

For Client 1, we utilize $420$ portal-venous phase abdominal CT images collected for preoperative planning in gastric surgery, so the stomach part is inflated. For the pancreas, we did not notice any particular abnormalities. The resolution of volumes are (0.58-0.98, 0.58-0.98, 0.16-1.0) in the voxel spacing (x, y, z) in millimeter. Only pancreas regions are manually annotated using semi-automated segmentation tools. We randomly split the data set into 252 training volumes, 84 validation volumes, and 84 testing volumes.

For Client 2's dataset, we collected $486$ contrast-enhanced abdominal CT images, where all volumes are from patients with pancreatic cancer. Among the whole dataset, the voxel spacing (x, y, z) in millimeter of 40 volumes are (0.68, 0.68, 1.0) and the rest $446$ volume are (0.68, 0.68, 5.0) in millimeter. The segmentation labels contain the normal part of the pancreas and the tumor of pancreatic cancer. All the labels are manually segmented by physicians. We split client 2's dataset into training, validation and testing sets randomly, the training set contains $286$ volumes, the validation set contains $100$ volumes and the testing set also contains $100$ volumes.

\subsection{Data Pre-Processing}
We re-sample the resolution of all volumes to isotropic spacing $1.0\mathrm{mm} \times 1.0\mathrm{mm} \times 1.0\mathrm{mm}$, and apply intensity clipping with minimum Hounsfield unit (HU) intensity $-200$ and maximum intensity $250$. Then we re-scale the value range to $[-1.0, 1.0]$.

\subsection{Neural Network Model}
We utilized the resulting model of coarse-to-fine network architecture search (C2FNAS)~\cite{Yu2019-qi}. The C2FNAS search algorithm performs a coarse-level search followed by a fine-level search to determine the optimal neural network architecture for 3D medical image segmentation. In the coarse-level search, C2FNAS searched for the topology of U-Net like models. In the fine-level search, C2FNAS searched for the optimal operations (including 2D convolution, 3D convolution, and pseudo-3D convolution) for each module from the previous search results. 

\begin{figure}
    \centering
    \includegraphics[width=1.0\textwidth]{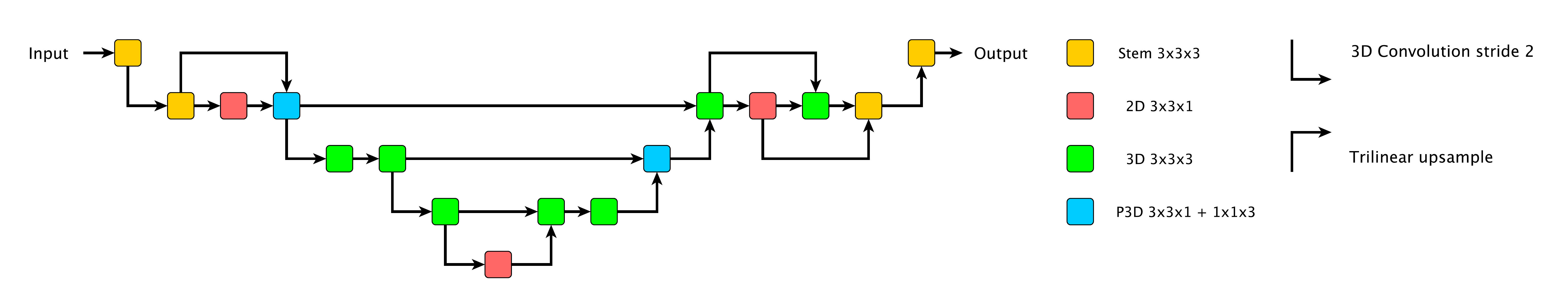}
    \caption{Model architecture of C2FNAS} 
    \label{fig:c2fnas}
\end{figure}

We add a VAE branch to the encoder endpoint of the C2FNAS model. The VAE branch shares encoder layers with C2FNAS and estimates the mean and standard deviation of encoded features for input image reconstruction. Two further losses, $L_{KL}$ and $L_{2}$, are introduced in~\cite{Myronenko2019-ye} are required for the VAE branch. $L_{KL}$ estimates the distance of mean vector and standard deviation from a Gaussian distribution, and the $L_{2}$ computes the distance of decoded volume and input volume in voxel level. VAE is capable of regularizing the shared encoder of the segmentation model.

Our implementation of VAE estimates the mean vector and the standard deviation vector by adding two separate dense layers with 128 output logits. In training, we construct the latent vector by adding mean vector and weighted standard deviation vector by random coefficients in normal distribution. In the validation and testing, we treat the mean vector as the latent vector. To reconstruct the input image, we add a dense layer to recover the shape of input features from the latent vector. With the recovered features, we use trilinear up-sampling and residual convolutional blocks to reconstruct the input image.

\begin{figure}
    \centering
    \includegraphics[width=0.8\textwidth]{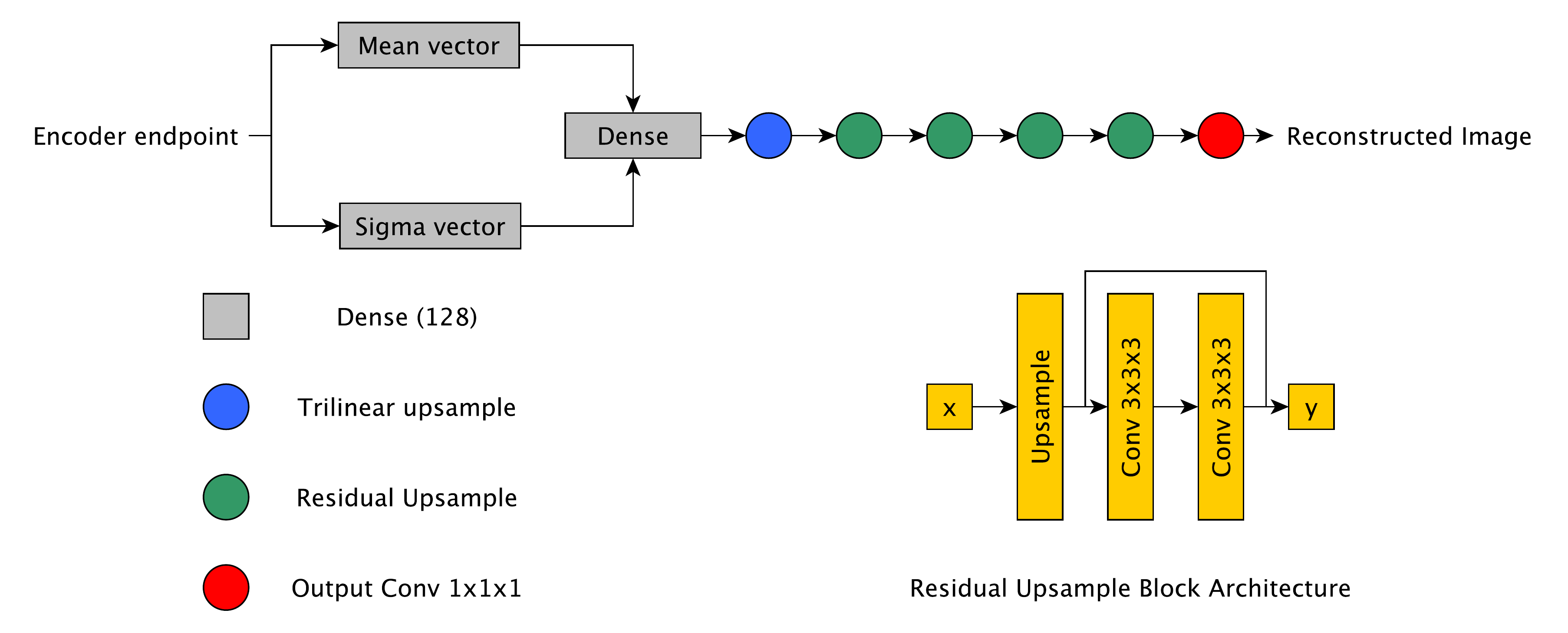}
    \caption{Model architecture of image reconstruction for variational auto-encoder.}
    \label{fig:vae}
\end{figure}

\subsection{Training Setup \& Implementation}
We use batch size 8 with 4 NVIDIA GPUs (Tesla V100 32GB for client 1 and Quadro RTX8000 for client 2) at each client in all our experiments, and the patches in each batch are randomly sampled and cropped from input volume. The sample rate of foreground patches and background patches are equal. The input patch size we use for training is $[96, 96, 96]$. We use Adam optimizer with learning rate ranged from $10^{-4}$ to $10^{-5}$, with cosine annealing learning rate scheduler. The loss for C2FNAS segmentation is Dice loss combined with categorical cross-entropy loss. In the setting with VAE regularization, we add VAE loss $L_{KL}$ and reconstruction loss $L_{2}$ to the total loss with constant coefficients $0.2$ and $0.3$, respectively.

Our implementation of the C2FNAS model is based on TensorFlow\footnote{\url{https://www.tensorflow.org/}}. Our FL experiments utilize the NVIDIA Clara Train SDK\footnote{\url{https://developer.nvidia.com/clara}} for model training and communication of weights between the server and clients.

\section{Experimental Results}
The experimental setups include standalone training on both clients and federated learning with two clients. In the standalone setting, both Client 1 (\textbf{C1}) and Client 2 (\textbf{C2}) train their local model independently with each client's own dataset, the resulting models are \textbf{C1\_baseline} and \textbf{C2\_baseline}. In the federated learning setup, we set up an aggregation server with no access to any dataset, two clients training on their local datasets sending gradients every ten epochs. The resulting models for federated learning are  \textbf{FL\_global}, \textbf{C1\_FL\_local} and \textbf{C2\_FL\_local}.

\begin{table}[tb]
\centering
\caption{Dice scores of pancreas and tumor. Data from C1 only have label for panaceas. FL improves the generalizability of model. }
\begin{tabular}{l|c|cc|c|c}
\hline
\multicolumn{1}{c|}{} & \textbf{C1} & \multicolumn{2}{c|}{\textbf{C2}} &  & \multicolumn{1}{c}{} \\ \cline{2-4}
\multicolumn{1}{c|}{\multirow{-2}{*}{\textbf{Dice coefficient}}} & \textbf{Pancreas} & \textbf{Pancreas} & \textbf{Tumor} & \multirow{-2}{*}{\textbf{Pancreas average}} & \multicolumn{1}{c}{\multirow{-2}{*}{\textbf{Average}}} \\ \hline
\textbf{C1\_baseline} & 81.5\% & 42.4\% & 0.0\% & 60.3\% & 30.1\% \\
\textbf{C2\_baseline} & 64.7\% & 65.4\% & \textbf{54.5\%} & 65.1\% & 59.8\% \\
\textbf{C1\_FL\_local} & 81.6\% & 65.2\% & 50.2\% & 72.7\% & 61.4\% \\
\textbf{C2\_FL\_local} & 81.6\% & {\color[HTML]{1D1C1D} \textbf{66.2\%}} & 52.6\% & \textbf{73.2\%} & \textbf{62.9\%} \\
\textbf{FL\_global} & \textbf{82.3\%} & {\color[HTML]{1D1C1D} 65.4\%} & 46.4\% & 73.1\% & 59.8\% \\ \hline
\textbf{Average} & 78.3\% & 60.7\% & 38.3\% & 68.7\% & 53.5\% \\ \hline
\end{tabular}
\label{tab:dice}
\end{table}

%
\begin{figure}[tb]
\centering
\begin{tabular}{ccc}
\subfloat[Ground truth]{\adjincludegraphics[valign=c,width=0.3\textwidth]{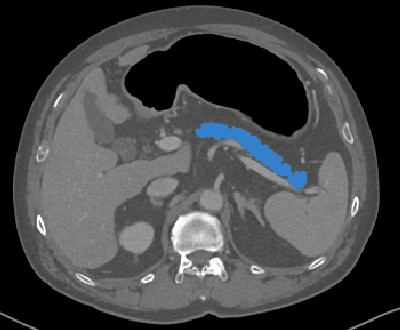}} &  
\subfloat[C1\_baseline]{\adjincludegraphics[valign=c,width=0.3\textwidth]{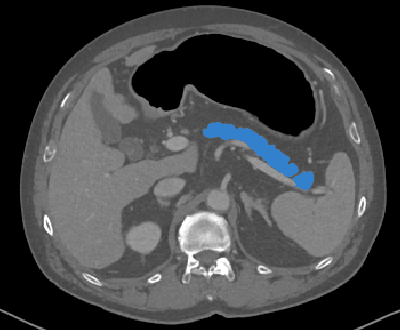}}&  
\subfloat[C2\_baseline]{\adjincludegraphics[valign=c,width=0.3\textwidth]{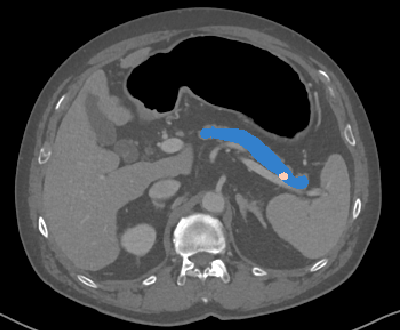}} \\	
\subfloat[FL\_global]{\adjincludegraphics[valign=c,width=0.3\textwidth]{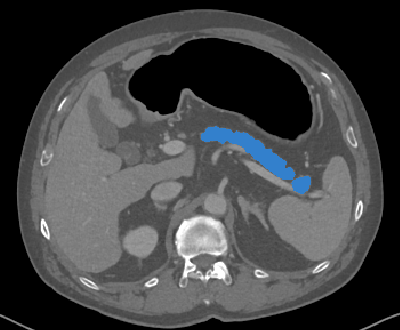}}&  
\subfloat[C1\_FL\_local]{\adjincludegraphics[valign=c,width=0.3\textwidth]{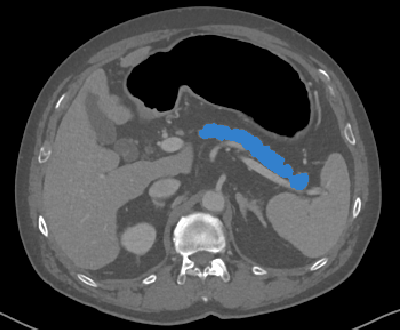}}&  
\subfloat[C2\_FL\_local]{\adjincludegraphics[valign=c,width=0.3\textwidth]{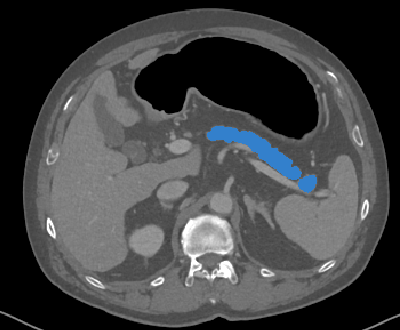}}
\end{tabular}
\caption{Comparison of segmentation results with the Client 1 (C1) dataset. Pancreas region in blue and yellow indicates the pancreas and tumor. Only pancreas regions are labeled in the Client 2 (C2) dataset.}
\label{fig:results-c1}
\end{figure}

\begin{figure}[tb]
    \centering
    \begin{tabular}{ccc}
        \subfloat[Ground truth]{\adjincludegraphics[valign=c,width=0.3\textwidth]{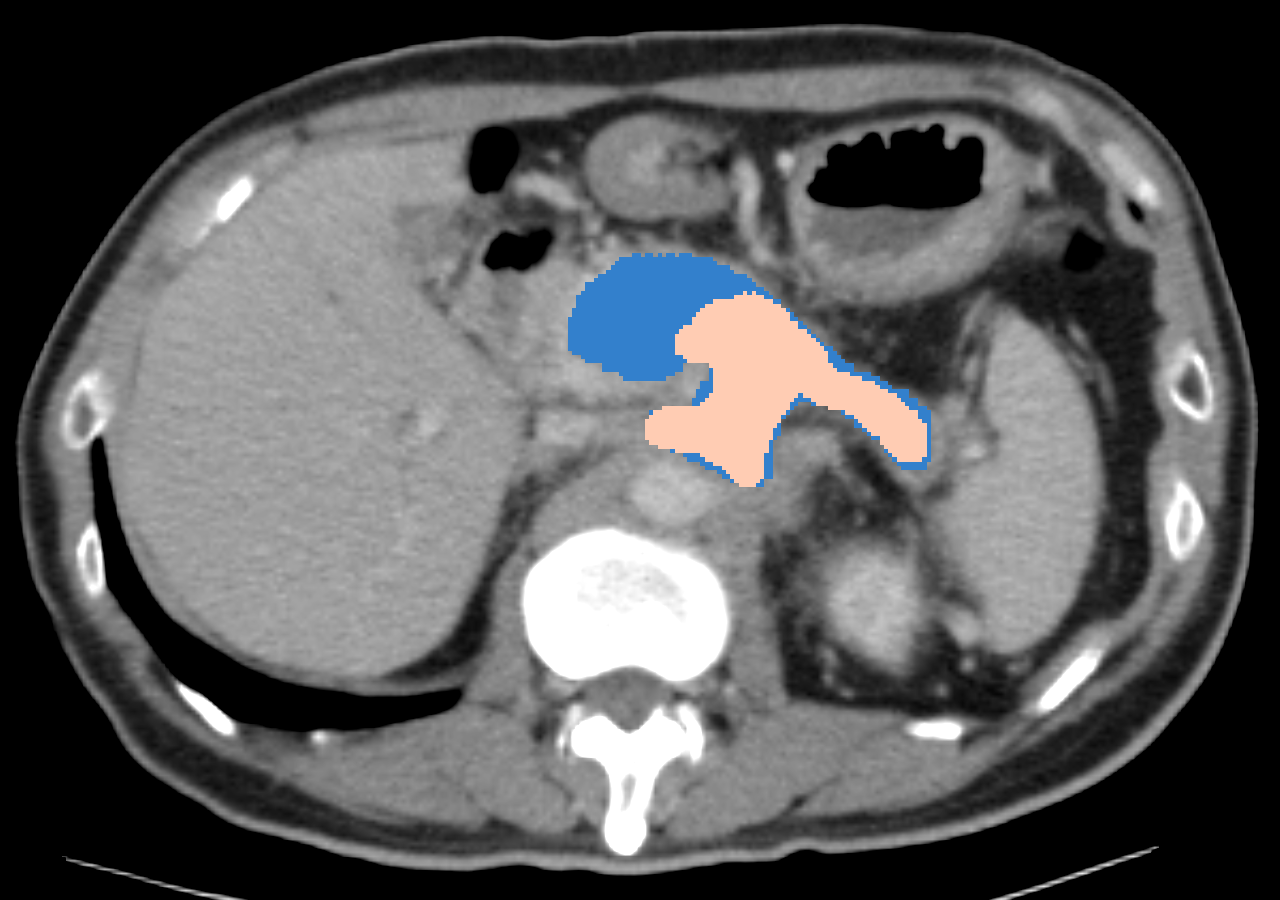}} &
        \subfloat[C1\_baseline]{\adjincludegraphics[valign=c,width=0.3\textwidth]{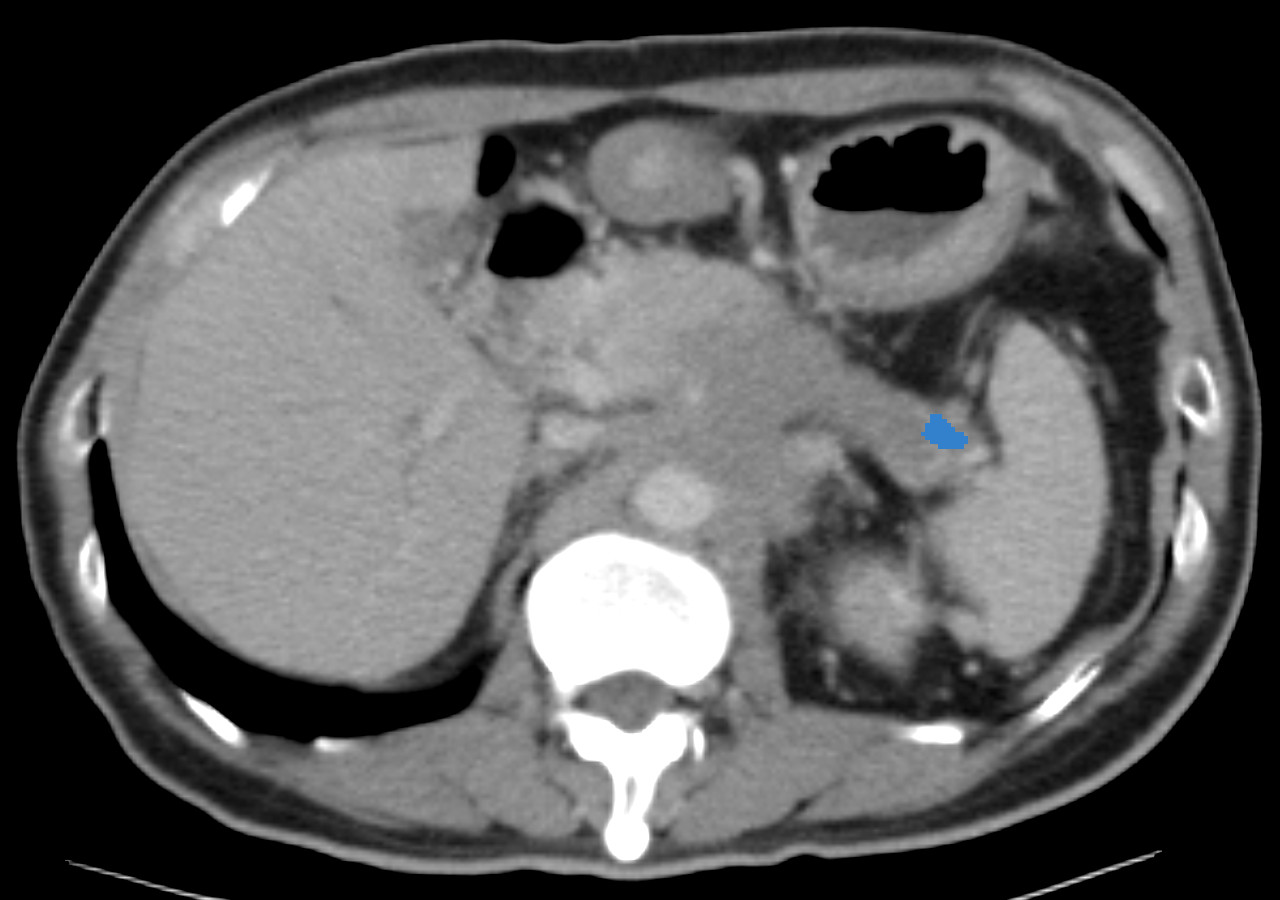}} &
        \subfloat[C2\_baseline]{\adjincludegraphics[valign=c,width=0.3\textwidth]{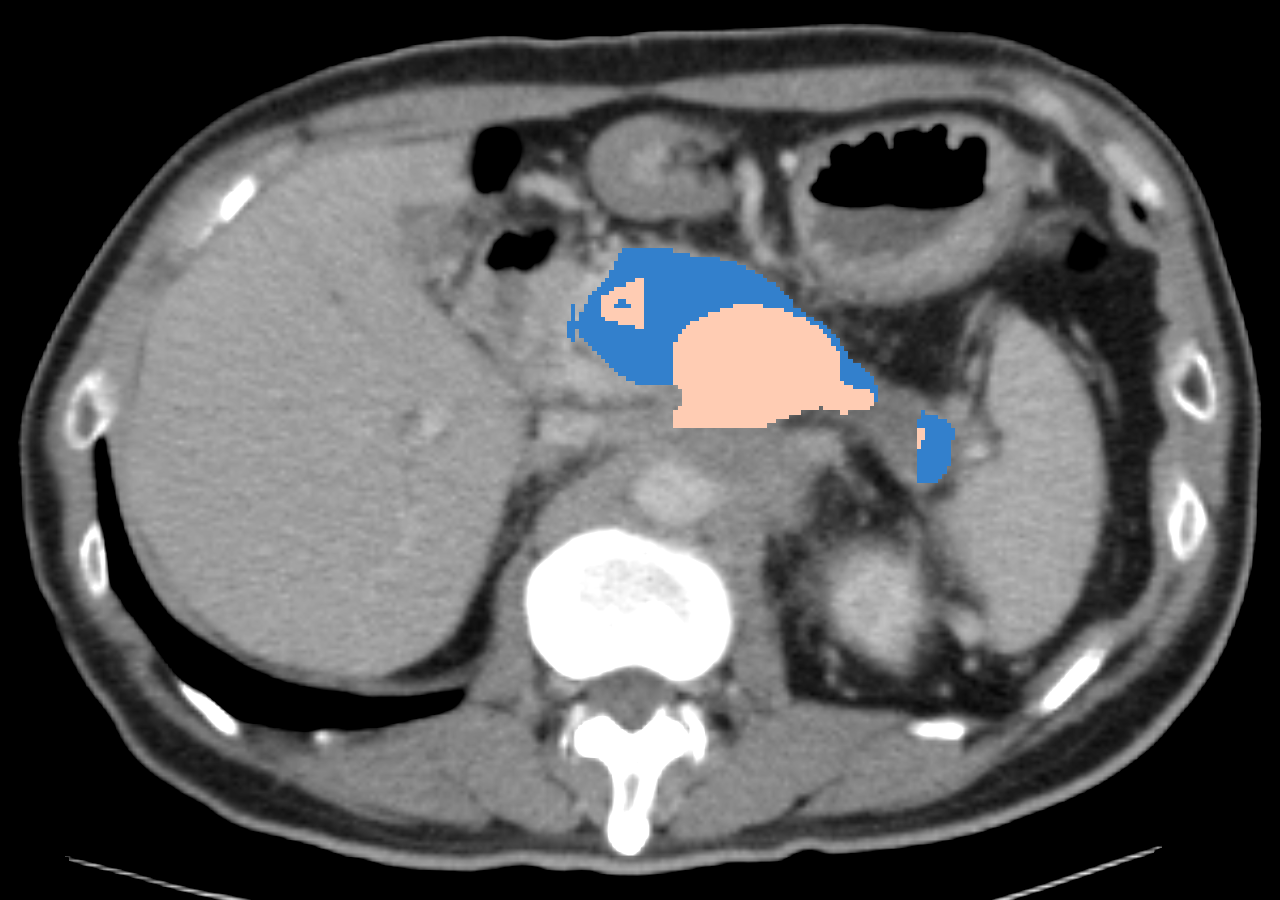}} \\
        \subfloat[FL\_global]{\adjincludegraphics[valign=c,width=0.3\textwidth]{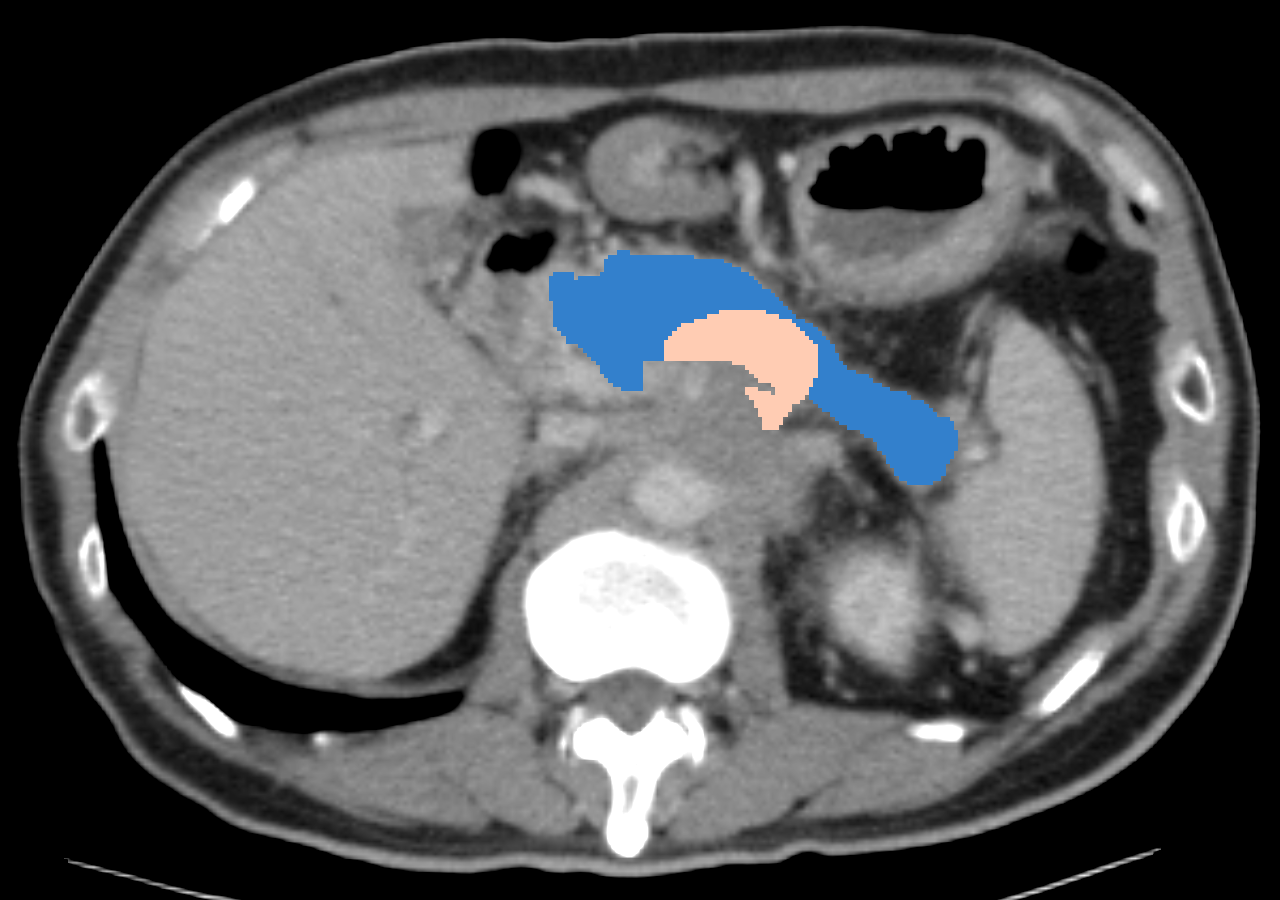}} &
        \subfloat[C1\_FL\_local]{\adjincludegraphics[valign=c,width=0.3\textwidth]{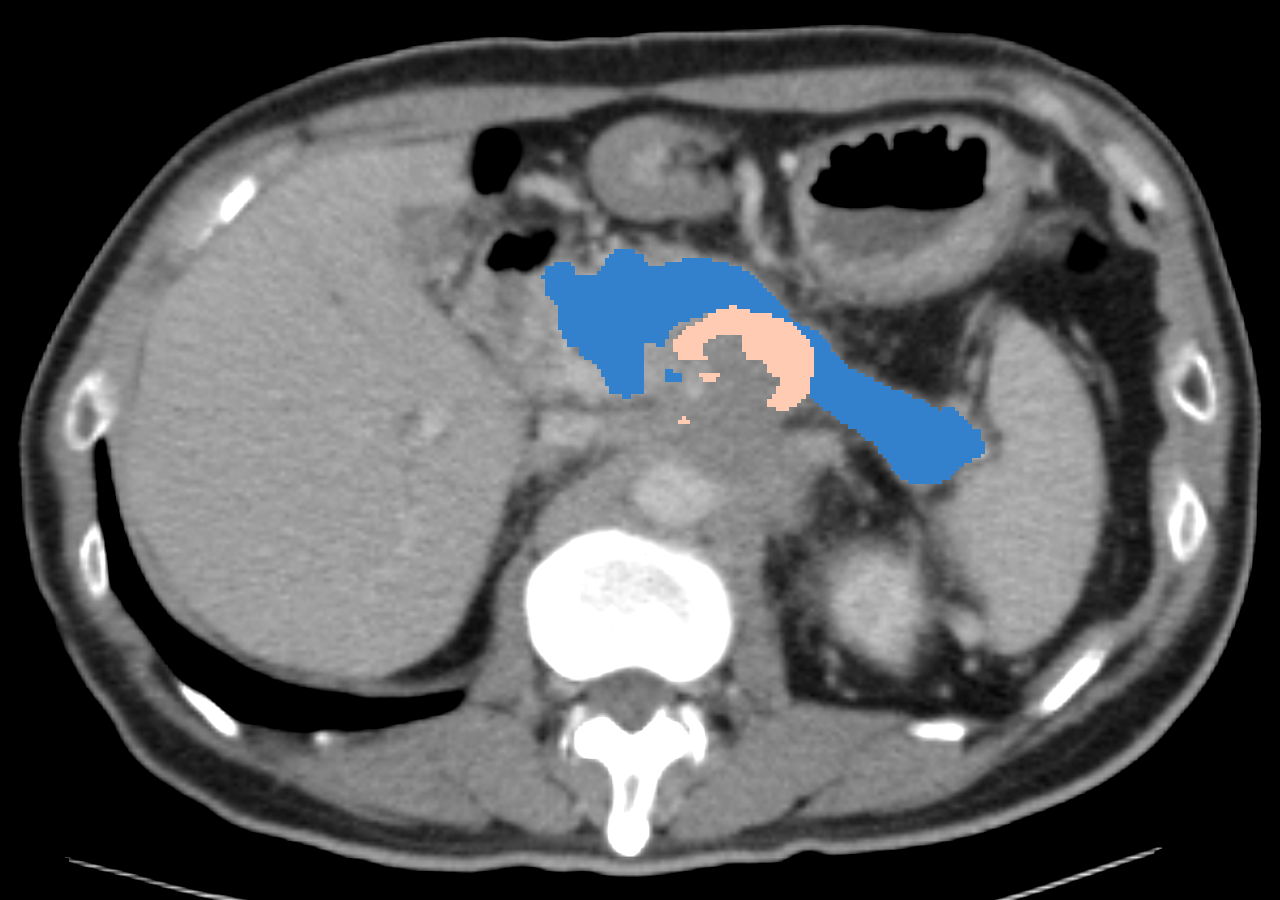}} &
        \subfloat[C2\_FL\_local]{\adjincludegraphics[valign=c,width=0.3\textwidth]{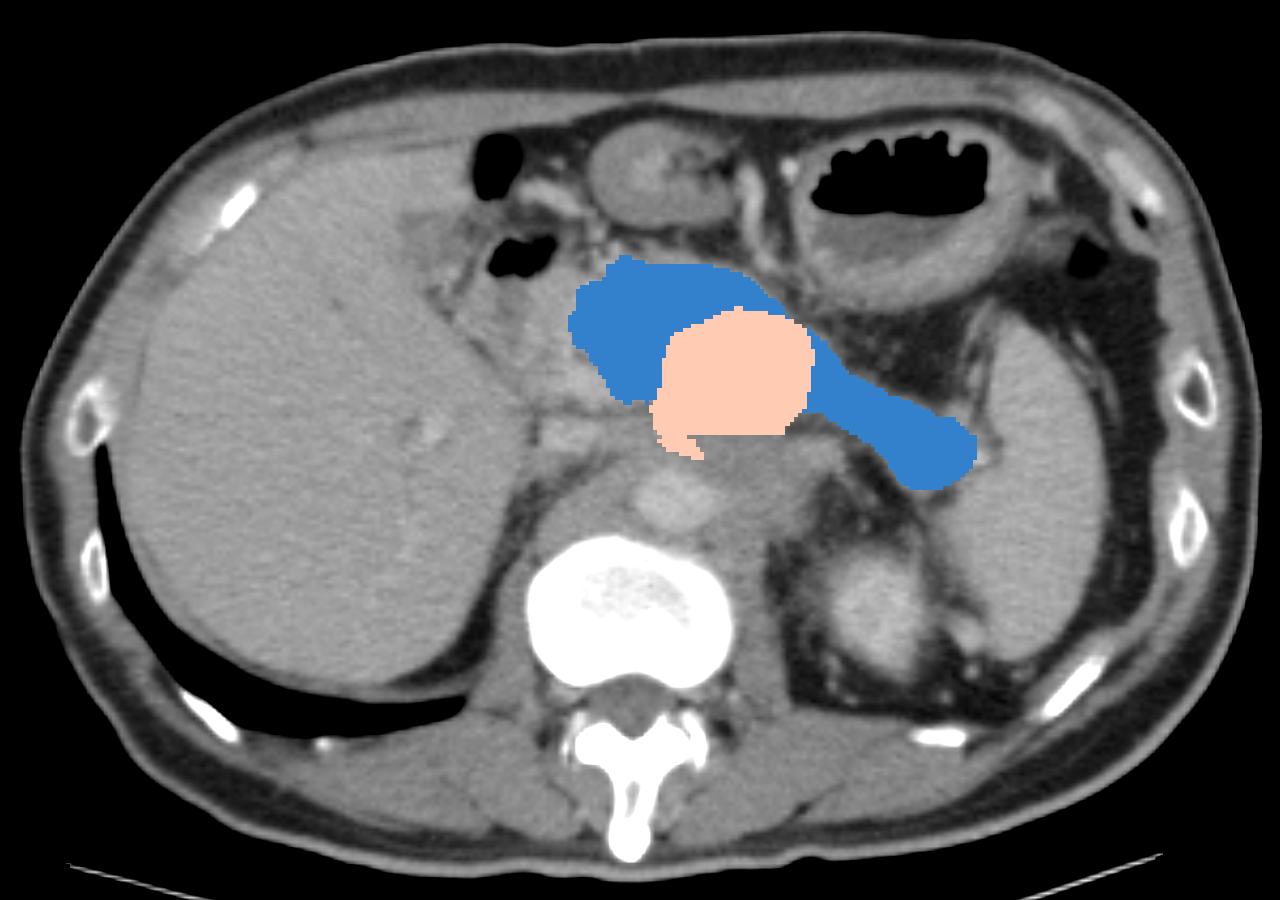}}
    \end{tabular} 
    \caption{Comparison of segmentation results with the Client 2 dataset. The pancreas part is labeled as blue and the tumor part is labeled as yellow.}
    \label{fig:resutls-c2}
\end{figure}

Table \ref{tab:dice} compare the standalone training models (C1\_baseline and C2\_baseline) and FL models (FL\_global, C1\_FL\_local, include C2\_FL\_local) for C1 dataset and C2 dataset. We have to mention that the C1 dataset only has pancreas label, whereas the C2 dataset is from pancreatic cancer patients, including pancreas and tumor label. For standalone models, the performance is not ideal when predicting on the other client's dataset. C2 tumor even get zero mean Dice socre on C1\_baseline model, because the C1 dataset does not include the tumor class. For FL models, the local model, both from C1 and C2, have great improvement on the other dataset. Segmentation performance for tumors on C1\_FL\_local model is comparable to a standalone model. Even the C1 dataset does not include tumor class. FL\_global model shows high generalizability on both C1 and C2 dataset.

Fig. \ref{fig:results-c1} shows the qualitative assessment on C1 dataset. When predicting with C2\_baseline model, a small region of the pancreas was misdetected as a pancreatic tumor, although CT volumes in the C1 dataset consist of healthy pancreas cases. The misdetection part disappeared after FL. FL\_global global model has the best performance on pancreas segmentation for the C1 dataset.

In Fig. \ref{fig:resutls-c2} we present a visualization of segmentation of one sample volume in the C2 test set. The prediction result of C1\_baseline model missed most areas of the pancreas and tumor. The prediction of C2\_baseline model is roughly in the correct area, but the shape of the tumor is incorrect and has a false positive of another tumor. The three federated learning models are doing better in detecting the area of the pancreas and the tumor. Although the tumor shape is still far from the ground truth in all predictions, the continuity of the area and the smoothness of the tumor boundary are significantly improved.

\section{Discussion}
In the standalone training setup, both C1\_baseline and C2\_baseline models perform well on their corresponding local test set. However, the testing results on the opposite test set have a significant performance drop. As the properties of the C1 dataset and C2 dataset are very different (healthy pancreas patients versus patients with pancreatic tumors), it is natural that the standalone models cannot generalize well to different data distribution.

In the federated learning setup, the performance of C1\_FL\_local model is slightly better than C1\_baseline in its own test set, and C1\_FL\_local model has a remarkable performance gain in the C2 test set, both the mean Dice score of pancreas and tumor on the C2 test set is comparable to the C2\_baseline model. For C2\_FL\_local model, the mean Dice score of the healthy part of the pancreas is slightly better than the C2 baseline model, and the mean Dice score of tumor part drops only moderately. The testing result of C2\_FL\_local model on the C1 dataset also has a substantial improvement from C2\_baseline model, and the performance is similar to C2\_FL\_local model. The FL\_global model can predict well for the pancreas for both test sets, but the prediction of tumors is notably lower than the other two local models. This drop is possibly caused by the lack of any validation or model selection procedure on the server-side. In the client training, we always keep the model with the highest local validation metrics, but on the server-side, the model aggregator only accepts gradients from the clients. The server cannot determine the quality of the model in our current training setting.
\section{Conclusions}
In this research, we conduct real-world federated learning to train neural networks between two institutes without the need for data sharing between the sites and despite inconsistent data collection criteria. The results suggest that the federated learning framework can deal with highly unbalanced data distributions between clients and can deliver more generalizable models than standalone training.
%
%
%
%
\bibliographystyle{splncs04}
\bibliography{references}
\end{document}